\documentclass[11pt]{article}
\usepackage{moriond,epsfig}

\def\mco{\multicolumn}

\def\be{\begin{equation}}
\def\ee{\end{equation}}
\def\bea{\begin{eqnarray}}
\def\eea{\end{eqnarray}}

\begin{document}
\vspace*{4cm}
\title{B-Hadron Lifetimes and $\Delta\Gamma$ at the Tevatron}

\author{ R.J. Lipton \\ (for the CDF and D\O\ Collaborations)}

\address{Fermilab, P.O. Box 500
Batavia, Il 60510 USA}

\maketitle\abstracts{We present new results on the lifetimes and widths of $B$ hadrons based 
on 300-450 pb$^{-1}$ of data collected by CDF and D\O\  at the Fermilab Tevatron. Lifetimes were measured in 
semileptonic decays as well as fully reconstructed hadronic modes.  A new measurement 
of the width difference between $B_s$ CP eigenstates, $\Delta\Gamma / \overline{\Gamma} $,  in $B_s$
 decays to $J/\psi \phi$ is also presented.}
 
\section{Introduction}
The long lifetimes of the $B$ mesons and baryons present unique opportunities 
for detailed examination of heavy quark interactions and weak decays.  
 CDF and D\O\ have access to the full range of $B$ hadrons and
 can measure lifetimes and lifetime ratios of a variety of states in the same 
 experiment. CDF and D\O\ have now analyzed 300-450 pb$^{-1}$ of $p\bar{p}$ data 
 at $1.96$ TeV and are beginning to provide precision measurements of lifetimes 
 of a variety of $B$ states. Both experiments have also made first measurements of 
 the lifetime difference between CP eigenstates, $\Delta\Gamma / \overline{\Gamma}$, in $B_s$ decays 
 to $J/\psi \phi$.
 
\section{Lifetimes}
Table~\ref{tab:oldlifetimes} provides a summary of Run II lifetime measurements from Tevatron experiments
prior to this conference. 
Lifetime ratios, which tend to have smaller experimental and 
theoretical errors than absolute measurements, are particularly useful in comparisons 
with heavy quark effective theory predictions~\cite{Gabbiani:2004tp}.
 These measurements represent the first lifetime 
data using large samples of fully reconstructed $B_s$ and $\Lambda_b$ decays.

D\O\ and CDF both use similar techniques for measuring lifetimes.  Decay vertices are 
reconstructed with tracks selected by kinematic and mass cuts.  The projection of the decay 
length is taken on the transverse plane ($L_{xy}$).  One then does a 
maximum likelihood fit to signal and background lifetimes (obtained from mass sidebands)
 and masses including resolution 
functions for momenta (for partially reconstructed vertices) and decay lengths as 
well as a scale factor for the decay length error.  

\begin{table} 
\caption{Tevatron Run II $B$ lifetime results prior to this conference.\label{tab:oldlifetimes}}
\vspace{0.4cm}
\begin{center}
\begin{tabular}{|l l l|}
\hline
 & D\O\ Results & CDF Results \\
\hline
\underline{Lifetimes} & &  \\
$B^0_s \to J/\psi \phi$ & $1.444 ^{+0.098}_{-0.090} \pm 0.020$ ps~\cite{Abazov:2004ce} & 
$1.369 \pm 0.100 ^{+0.008}_{-0.010} $ ps~\cite{CDF7409} \\
$\Lambda_b \to J/\psi \Lambda$ & $1.22 ^{+0.22}_{-0.18} \pm 0.020$ ps~\cite{Abazov:2004bn} & 
$1.25 \pm 0.26 \pm 0.10 $ ps~\cite{Paulini:2004rg} \\
$B_c \to J/\psi \mu $X & $0.448 ^{+0.123}_{-0.06} \pm 0.121$ ps~\cite{D04539} & \\ & & \\
\underline{Lifetime Ratios} & &  \\
$\tau({B^0_s}) / \tau({B^0})$ & $0.980 ^{+0.075}_{-0.070} \pm 0.04$~\cite{Abazov:2004ce} & \\
$\tau({\Lambda_b}) / \tau({B^0}) $ & $ 0.87 ^{+0.17}_{-0.14} \pm 0.03$~\cite{Abazov:2004bn} & \\
$\tau({B^+}) / \tau(B^0)$ & $1.080 \pm 0.016 \pm 0.014$~\cite{Abazov:2004sa} & 
$1.119 \pm 0.046 \pm 0.014$~\cite{CDF6387} \\
\hline
\end{tabular}
\end{center}
\end{table}

\subsection{Lifetimes in Semileptonic Decays}\label{subsec:cdfsemi}
Semileptonic decays typically  provide large event samples with degraded momentum resolution due to the 
missing neutrino.  D\O\ relies on it's muon trigger while CDF uses both muon and electron decay modes. The 
$B$ decay parent is implied by reconstruction the charmed daughters.  Corrections 
 must be applied for the relative branching ratios of the $B$ parents into the observed states. 
The composition of the sample must be well understood since the component decay modes have 
different effective momentum resolution functions (\textit{K} factors).

CDF uses both electron and muon triggers in 260 pb$^{-1}$ of data.  This sample has the advantage 
relative to the hadronic sample discussed in section 2.2 of having no lifetime-dependent trigger bias. 
The reconstructed final states are $D^*(e~or \mu) \nu X$ , or $D(e~or \mu)X$ with $D\to K^-\pi^+$. 
A simultaneous fit is made 
to both lifetime and mass distributions using the relative 
fractions of charged and neutral $B$ mesons 
decaying to $D$ and $D^*$ and Monte Carlo efficiency for the reconstruction of the $D^*$ pion.  
The results~\cite{CDF7514} are shown in Table~\ref{tab:leptonic}. The systematic uncertainty is dominated 
by uncertainties in the sample composition and estimates of the prompt charm background contributions.

\begin{table}
\caption{$B$ lifetimes measured in semi-leptonic modes.\label{tab:leptonic}}
\vspace{0.4cm}
\begin{center}
\begin{tabular}{|l l l|}
\hline
State & Decay Mode & Lifetime or Ratio \\
\hline
\mco{3}{|c|}{\underline{CDF}} \\
$B^-$ & $l \nu D^0 X$, $D^0 \to K \pi$   & $1.653 \pm 0.029 ^{+0.033}_{-0.031}$ ps \\
$B^0$ & $l \nu D^* X$, $D^* \to D^0 \pi$ & $1.473 \pm 0.036 \pm 0.054$ ps \\
$\tau(B^+) / \tau(B^0)$ & & $1.123 \pm 0.040 ^{+0.041}_{-0.039}$ \\
\hline
\mco{3}{|c|}{\underline{D0}} \\
$B_s$ & $D_s \mu \nu X$, $D_s \to \phi \pi$ & $1.420 \pm 0.043 \pm 0.057$ \\
\hline
\end{tabular}
\end{center}
\end{table}

\begin{figure}
\begin{tabular} {c c}
(a) \epsfig{figure=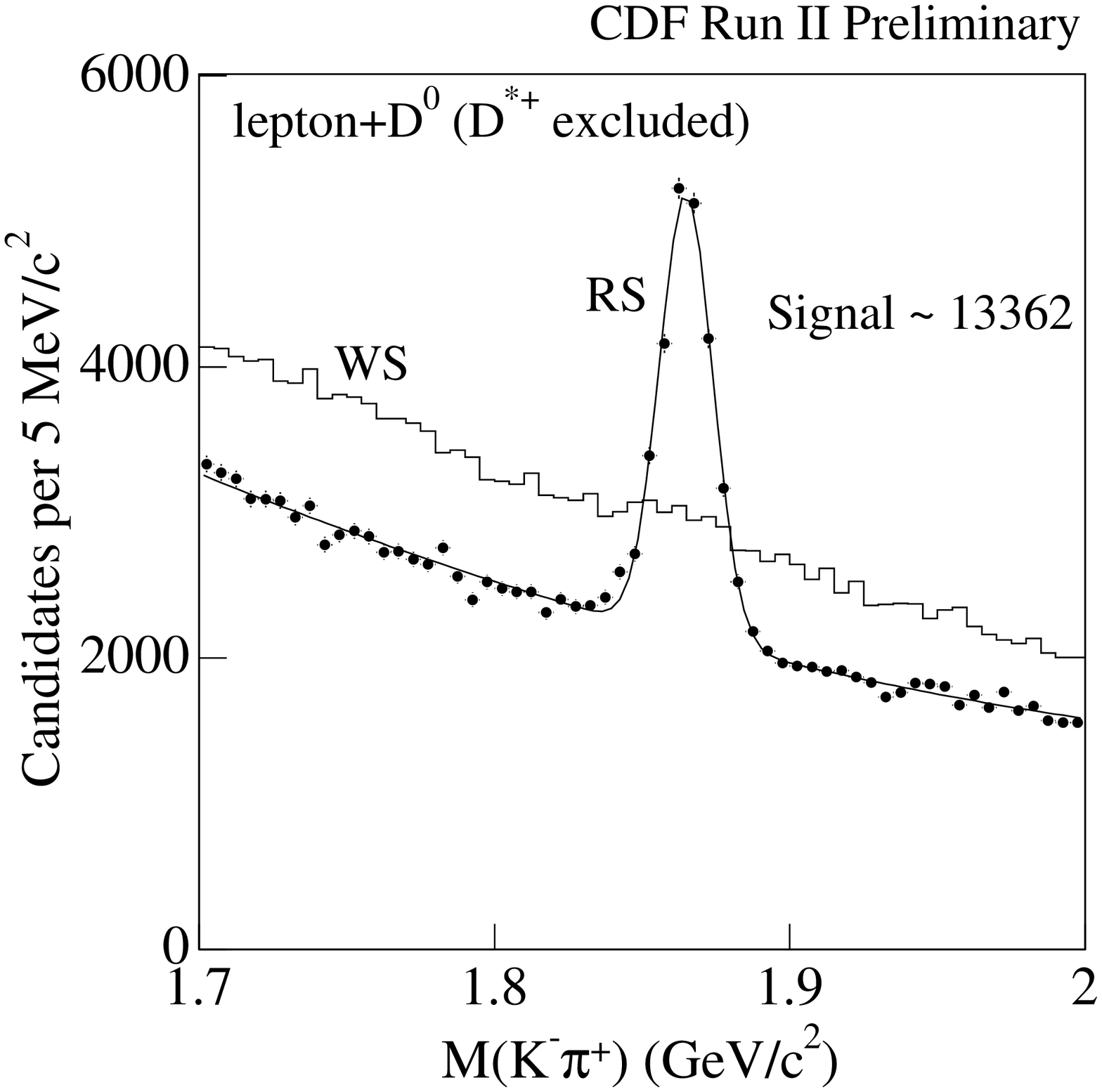 ,height=2.5in}
(b) \epsfig{figure=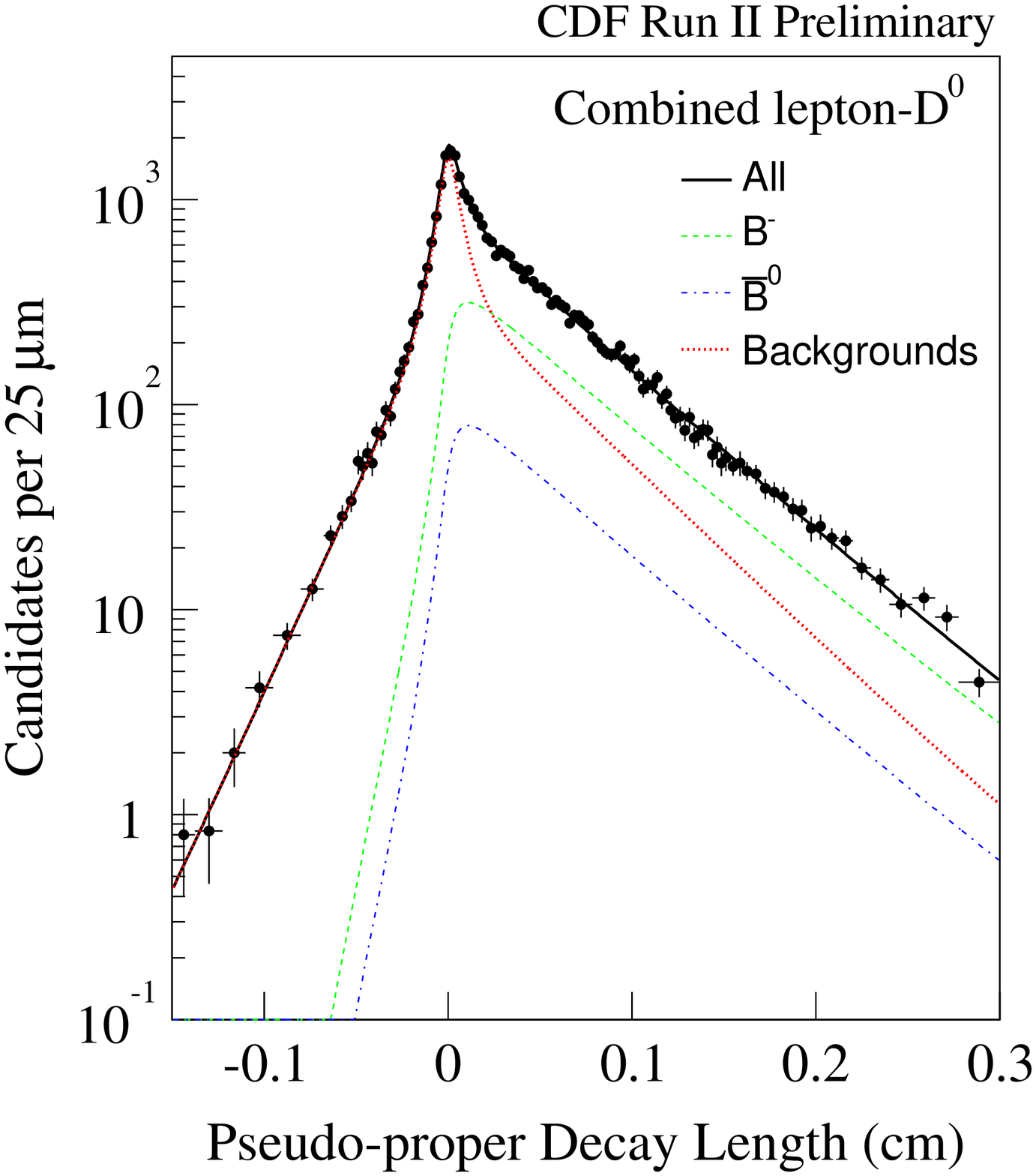 ,height=2.5in} \\
(c) \epsfig{figure=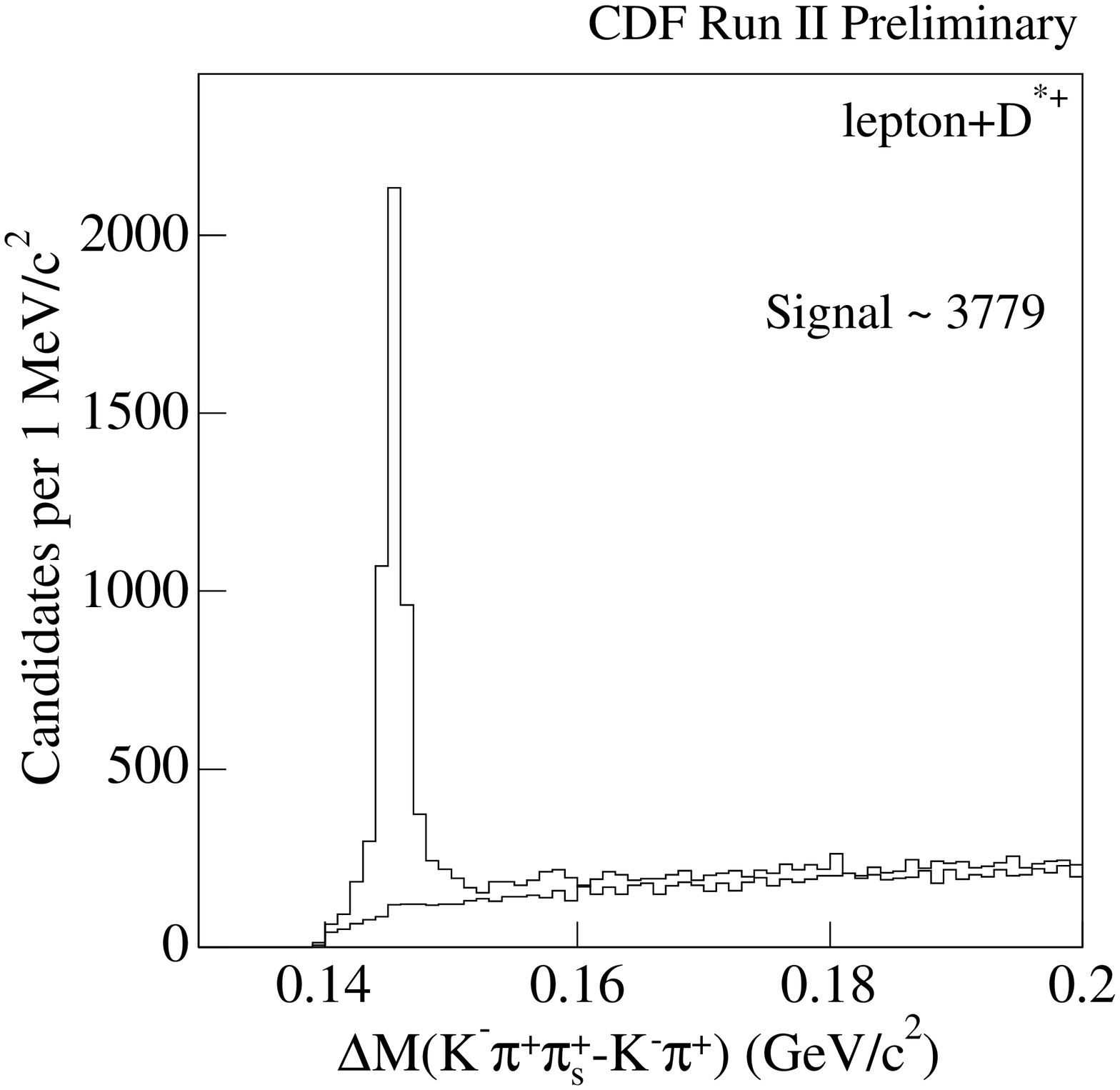 ,height=2.5in}
(d) \epsfig{figure=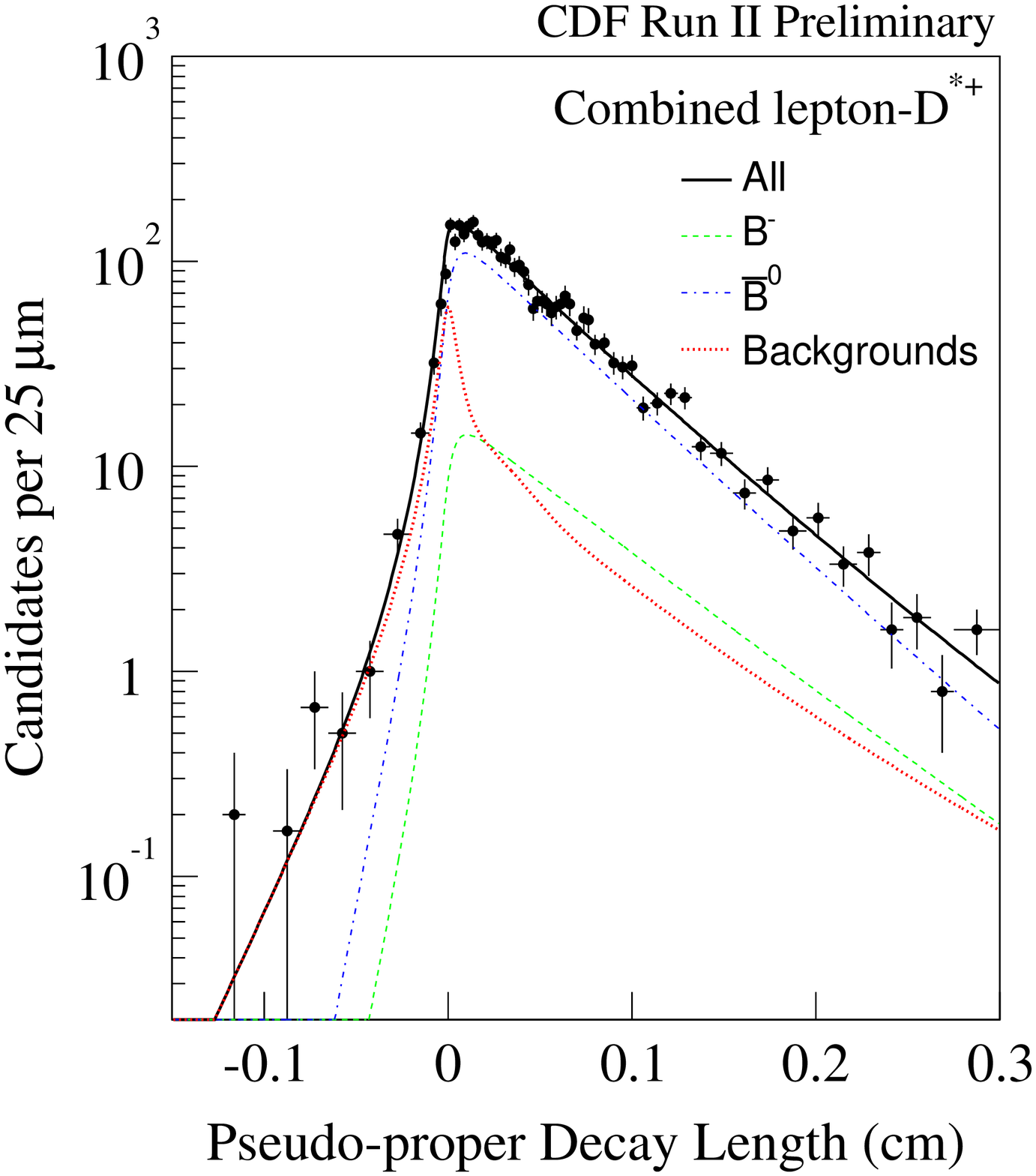 ,height=2.5in}
\end{tabular}
\caption{Mass and lifetime distributions for the $B \to D l \nu X$ (a,b)
and $B \to D^* l \nu X$ (c,d) samples. Lines are results of the maximum 
likelihood fit. \label{fig:cdfsemi}}
\end{figure}

\begin{figure}
\begin{tabular} {c c}
(a) \epsfig{figure=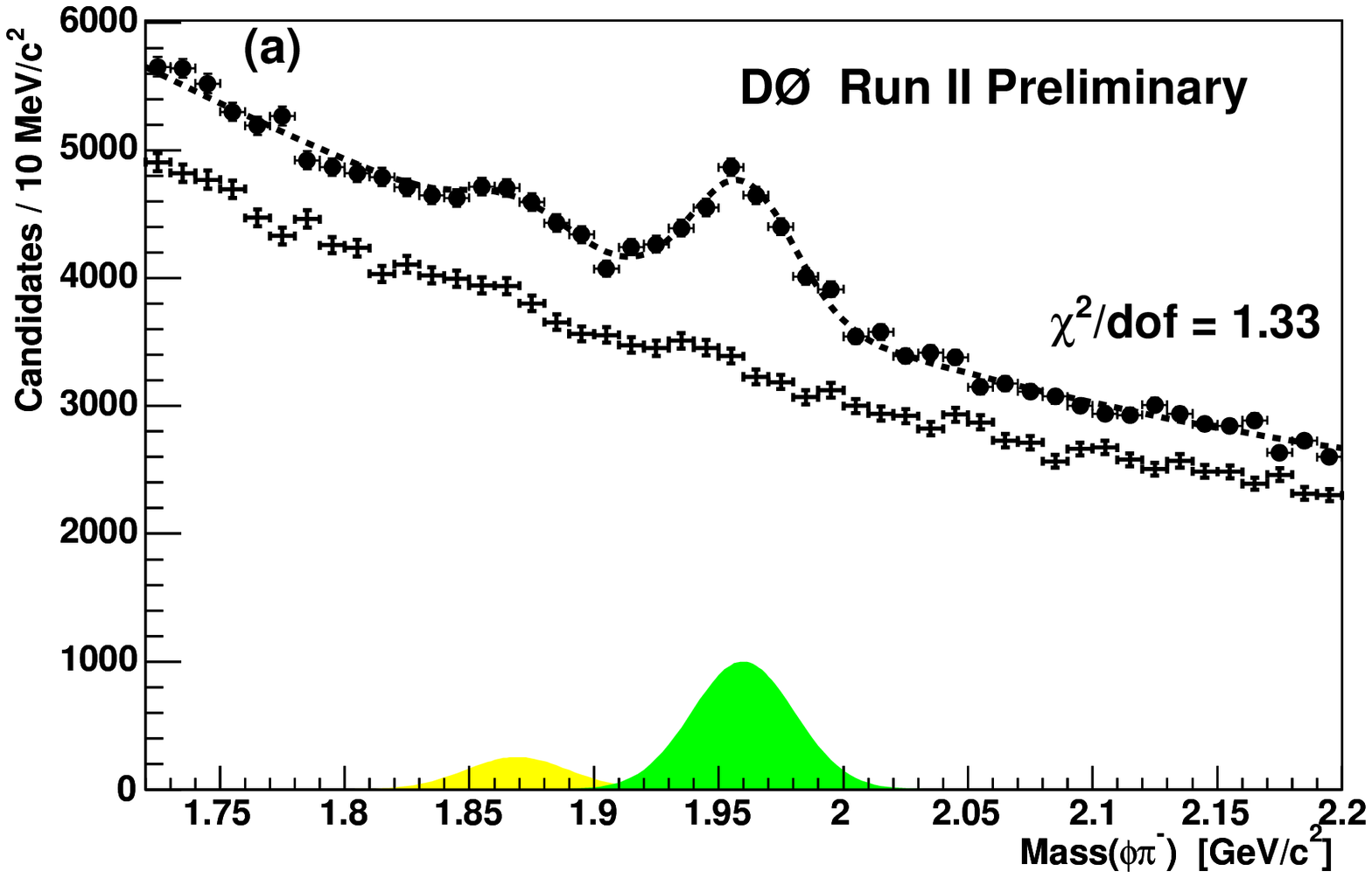 ,height=2in} &
(b) \epsfig{figure=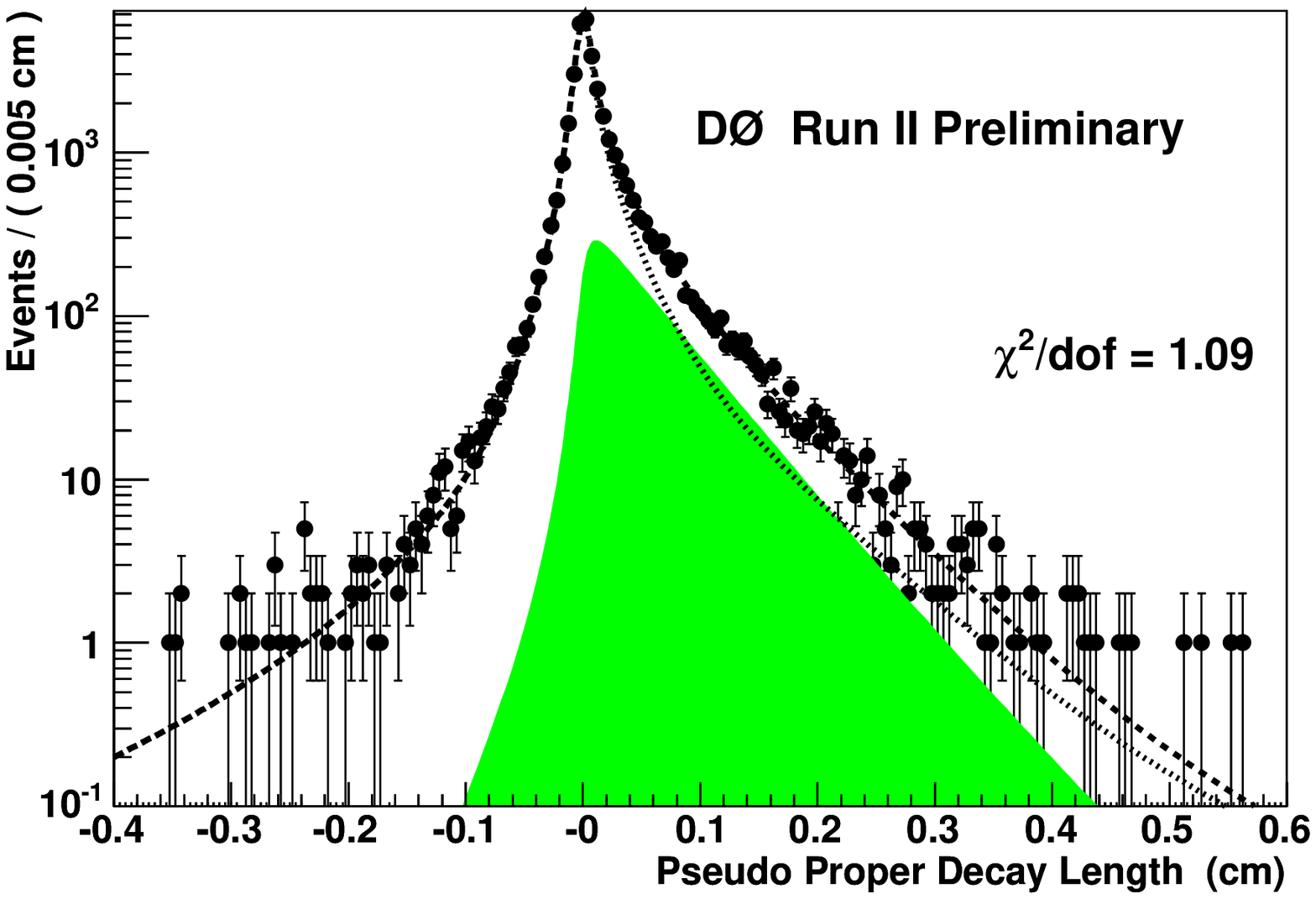 ,height=2in}
\end{tabular}
\caption{Mass (a) and lifetime (b) distributions for $B_{s}\to D_s \mu \nu X$ from D\O\ 
with results of the likelihood fit overlaid. \label{fig:d0semi}}
\end{figure}

D\O\ has a new measurement of the $B_s$ lifetime in the decay $B_{s}\to D_s \mu \nu X$, $D_s \to \phi \pi$
 in 400 pb$^{-1}$ of data. The fit includes calculated contamination from $B^0$, $B^+$ and 
 a momentum distribution which also includes prompt $D^*_s$ contributions.  The background at large negative lifetime 
 comes from directly produced charm pairs with misreconstructed vertices. The mass and lifetime distributions 
 are shown in Fig.~\ref{fig:d0semi}. The result~\cite{D04729}, included in Table~\ref{tab:leptonic}, is
 $\tau (B_s) = 1.420 \pm 0.043 \pm 0.057$.  Systematics are dominated by the uncertainty in the extrapolation 
 of sideband lifetimes to the $B_s$ signal region.

\subsection{Lifetimes in Hadronic Modes}\label{subsec:cdfhadr}
\begin{figure}
\begin{tabular} {c c}
(a) \epsfig{figure=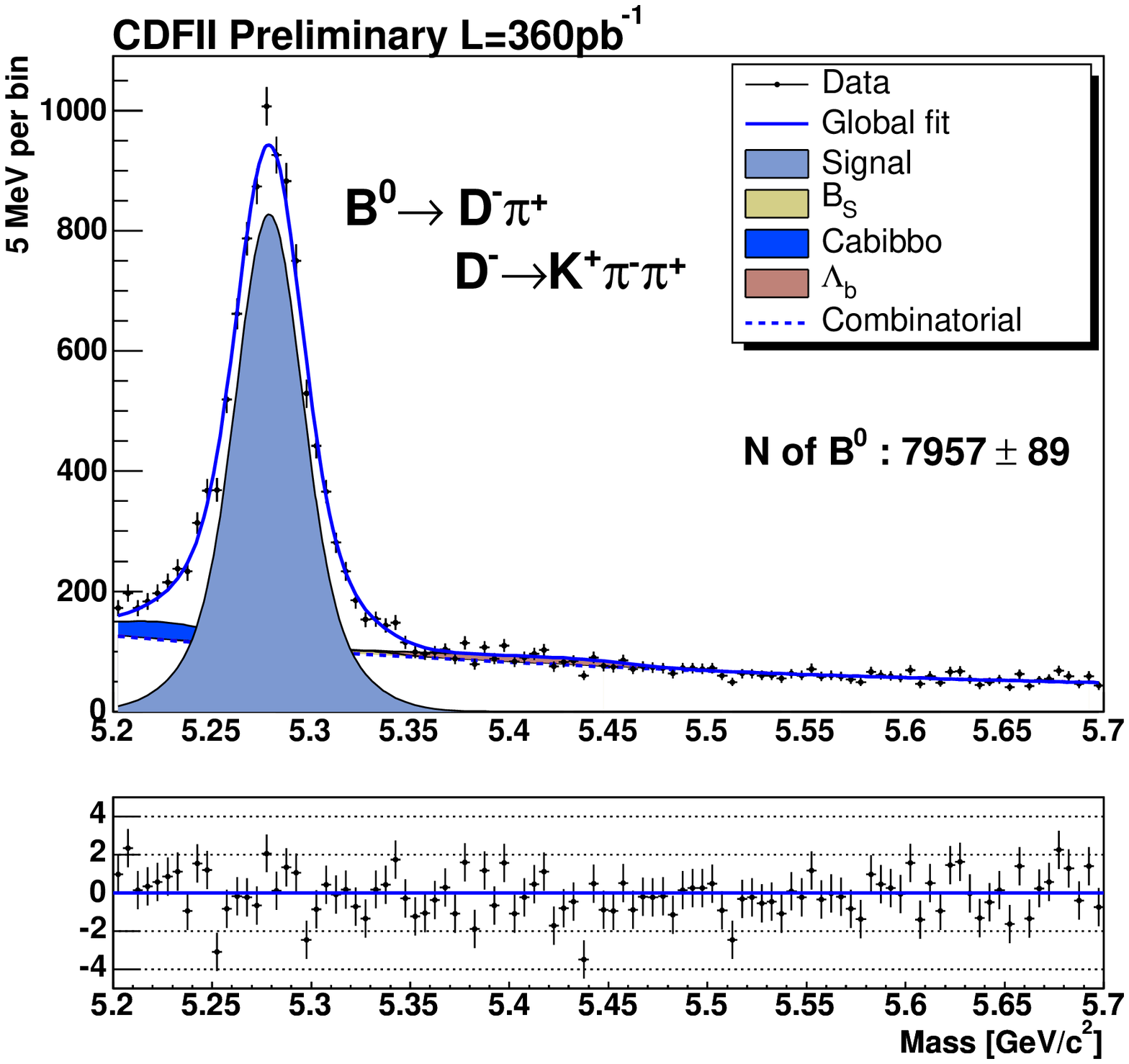 ,height=2in} &
(b) \epsfig{figure=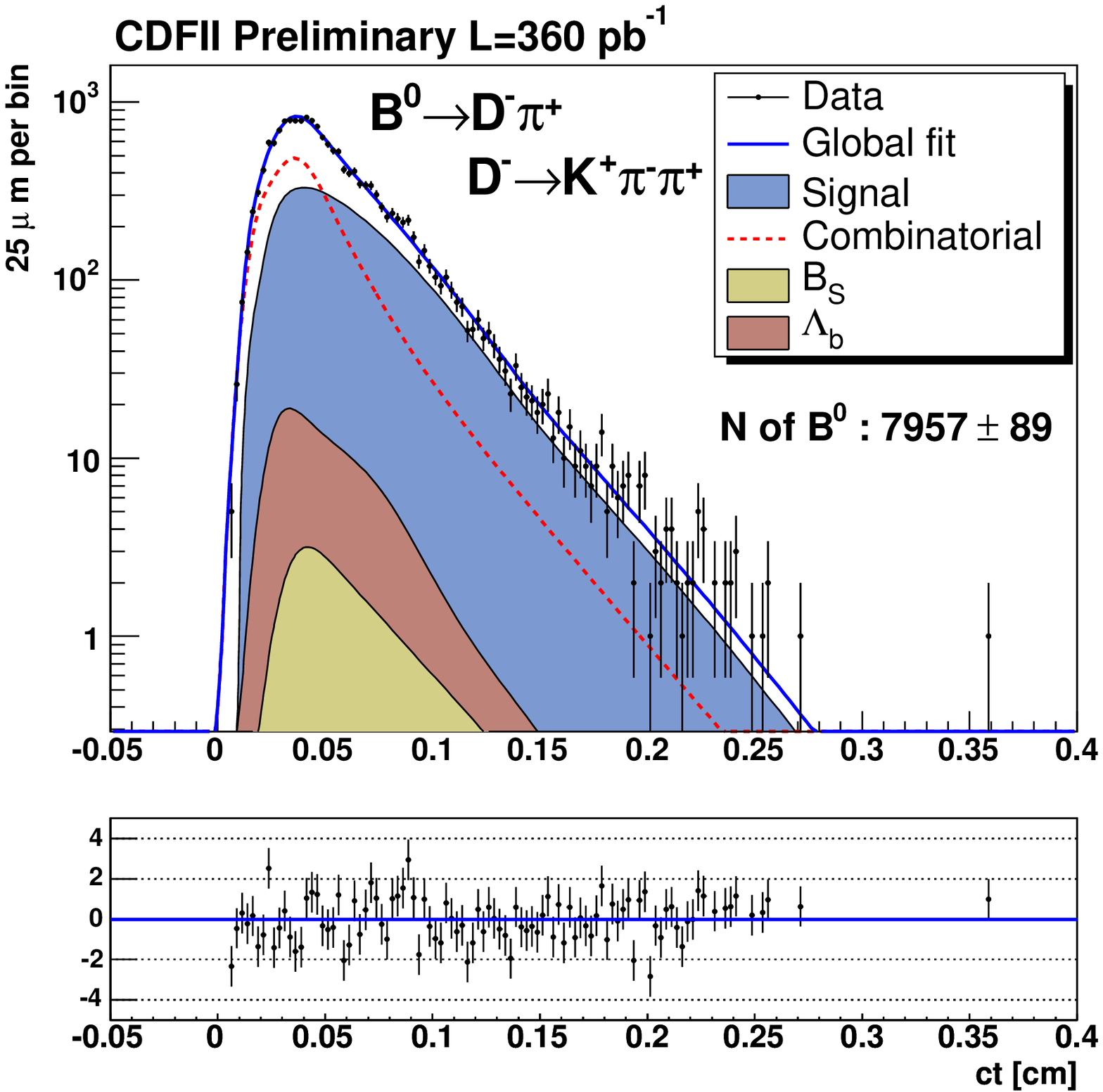 ,height=2in}
\end{tabular}
\caption{Mass (a) and lifetime (b) distributions for the $B^0 \to $ D$^- \pi^+$
mode in events triggered by the silicon vertex trigger. \label{fig:cdfhadlife}}
\end{figure}

CDF has presented first lifetime results in hadronic decay modes in data utilizing their silicon vertex trigger
(SVT) in 360 pb$^{-1}$ of data.  This level 2 trigger selects events including tracks with significant impact parameters.  The selection introduces a bias in the reconstructed lifetime distribution.  
As can be seen in Fig.~\ref{fig:cdfhadlife} the lifetime distributions lack the usual prompt peak.
This is corrected using a decay length and final state dependent 
efficiency generated by Monte Carlo and checked using charged $B^-\to J/\psi K^-$ events 
triggered by dimuons from the $J/ \psi$ as well as the SVT. Typical mass and 
decay length distributions are shown in Figure~\ref{fig:cdfhadlife}. A mass-only fit using a 
wide mass window is first used to establish the contributions of various backgrounds.   
The combined mass and lifetime fit is then performed in a narrower mass range. The background 
proper time is modeled using the high mass sideband.  The overall systematic error in these 
measurements is estimated to be less than 5 $\mu$m. 
Results~\cite{CDF7386} for the five modes studied are shown in Table~\ref{tab:CDFhadronic}.

\begin{table}
\caption{$B$ lifetimes measured in hadronic modes triggered by the SVT.\label{tab:CDFhadronic}}
\vspace{0.4cm}
\begin{center}
\begin{tabular}{|l l l|}
\hline
State & Decay Mode & Lifetime or Ratio \\
\hline
$B^0$ & $D^- \pi^+$, $D^- \pi^+ \pi^- \pi^+ $  & $ 1.51 \pm 0.02 \pm 0.01 $ ps  \\
$B^+$ & $D^0 \pi^+$  & $1.66 \pm 0.03 \pm 0.01$ ps  \\
$B^0_s$ & $D_s^- \pi^+$, $D_s^- \pi^+ \pi^- \pi^+$ & $1.60 \pm 0.10 \pm 0.02$ ps  \\
$\tau(B^+) / \tau(B^0)$ &  & $ 1.10 \pm 0.02 \pm 0.01 $ \\
$\tau(B^0_s) / \tau(B^0)$ & & $ 1.06 \pm 0.07 \pm 0.01 $ \\
\hline
\end{tabular}
\end{center}
\end{table}

\begin{figure}
\begin{center}
\epsfig{figure=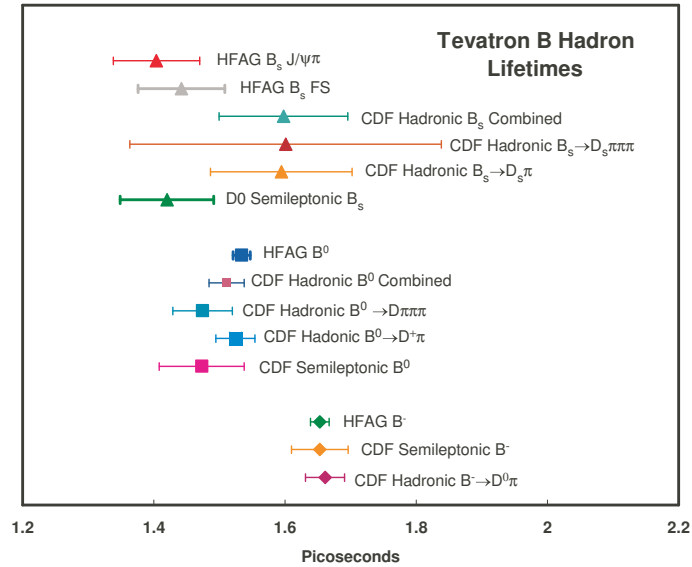 ,height=3in}
\end{center}
\caption{Summary of new lifetime measurements from CDF and D\O\ presented at this conference
along with HFAG world average values. \label{fig:lifesum}}
\end{figure}

Figure \ref{fig:lifesum} summarizes the new lifetime results from CDF and D\O\ for this conference along 
with the summer 2004 world averages compiled by the Heavy Flavor Averaging Group~\cite{Alexander:2005cx}. 
 In many cases the individual 
lifetimes measured by Tevatron experiments are competitive with current world averages. Much of 
the upcoming work will concentrate on reducing systematics, which are already small in the ratio 
measurements, in the absolute lifetimes.  

\section{ Lifetime Difference in the $B_s$ System, $\Delta \overline{\Gamma} $ }
Mass eigenstates in the $B_s$ system, $B_H$ and $B_L$, are related to the flavor eigenstates, 
$B^0_s$, $\overline{B}^0_s$ by:
\begin{eqnarray*}
B_H & = & p \mid B^0_s \rangle + q \mid \overline{B}^0_s \rangle , \\
B_L & = &  p \mid B^0_s \rangle - q \mid \overline{B}^0_s \rangle , \\
\Delta M  & = & M_H - M_L ,  \\
\Delta \Gamma & = & \Gamma_L - \Gamma_H , \overline{\Gamma} = \frac{\Gamma_L + \Gamma_H}{2}.
\end{eqnarray*}
Within the standard model $B_H$ and $B_L$ are expected 
to be approximate CP eigenstates ($p \approx q$) with distinct 
lifetimes.  The long-lived, heavy state is primarily CP even. 
The lifetime difference between these states, $\Delta \Gamma$, is governed by real intermediate 
states and has been calculated using 
the Heavy Quark Expansion,  $ \Delta \Gamma / \overline{\Gamma} = 0.12 \pm 0.05$ ~\cite{Lenz:2004nx}.  
CDF has reported a value of $ \Delta \Gamma / \overline{\Gamma} = 0.65 \pm 0.45$ in the summer of 2004~\cite{Acosta:2004gt}.
Such a large central value, if confirmed, would be difficult to explain 
theoretically since new physics effects would generally tend to lower $\Delta \Gamma~$\cite{Dunietz:2000cr}.

The CP content of decays of the $B_s$ to J/$\psi $ ( $\to \mu \mu$ )$ \phi $($ \to K K$) 
can be analyzed by a study 
of the angular distributions of the decay products.  The full angular distribution~\cite{Dighe:1995pd}
involves three angles ($\theta , \phi , \psi$)  and 
a fit provides the CP even (A$_0$, A$_\|$) and CP odd (A$_\bot$) amplitudes as well as a phase. 
The CDF result fit to all three angles and provided values for all amplitudes as well as the strong phase.
 
\begin{figure}
\begin{tabular} {c c}
(a) \epsfig{figure=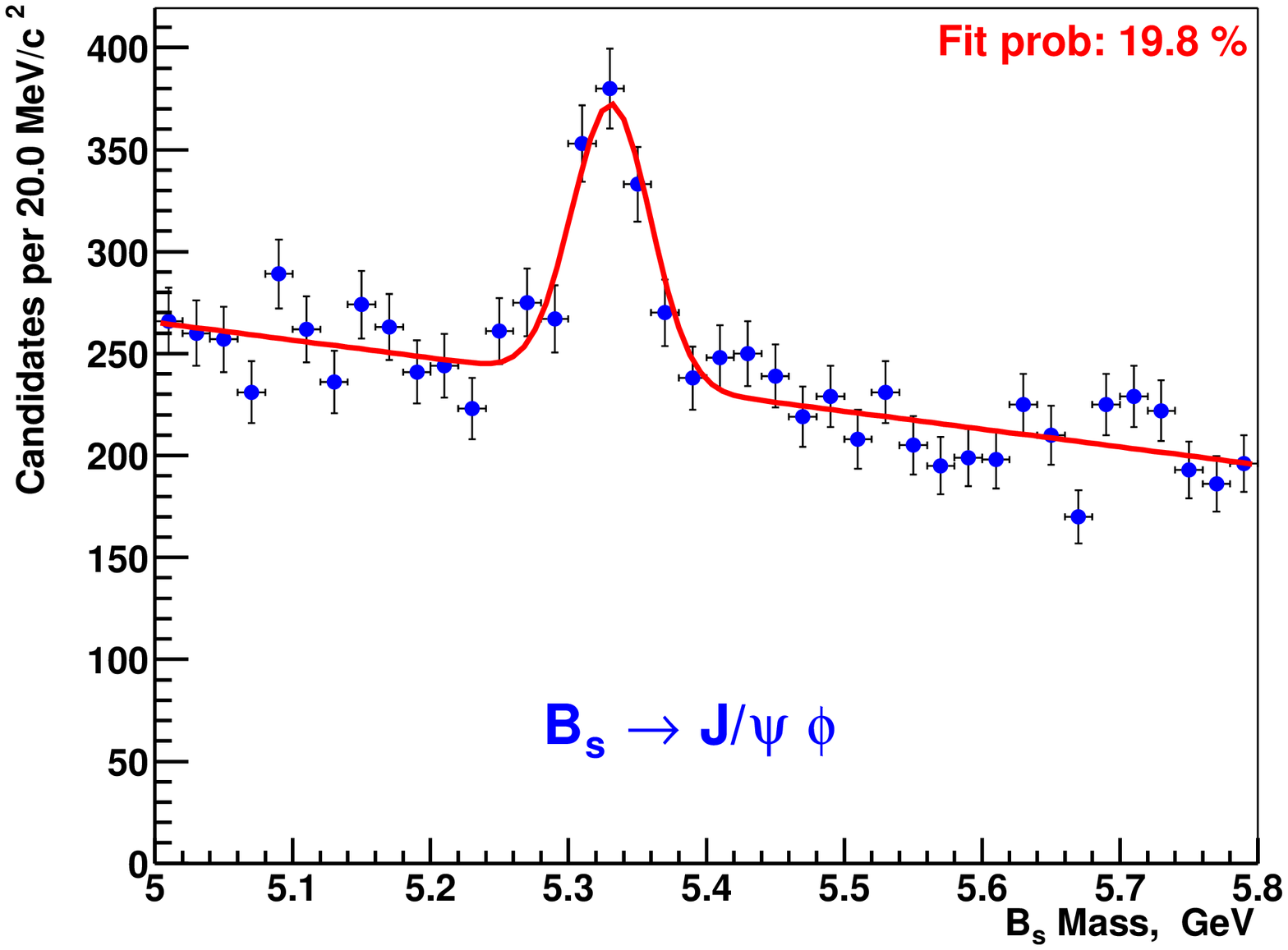 ,height=2in} &
(b) \epsfig{figure=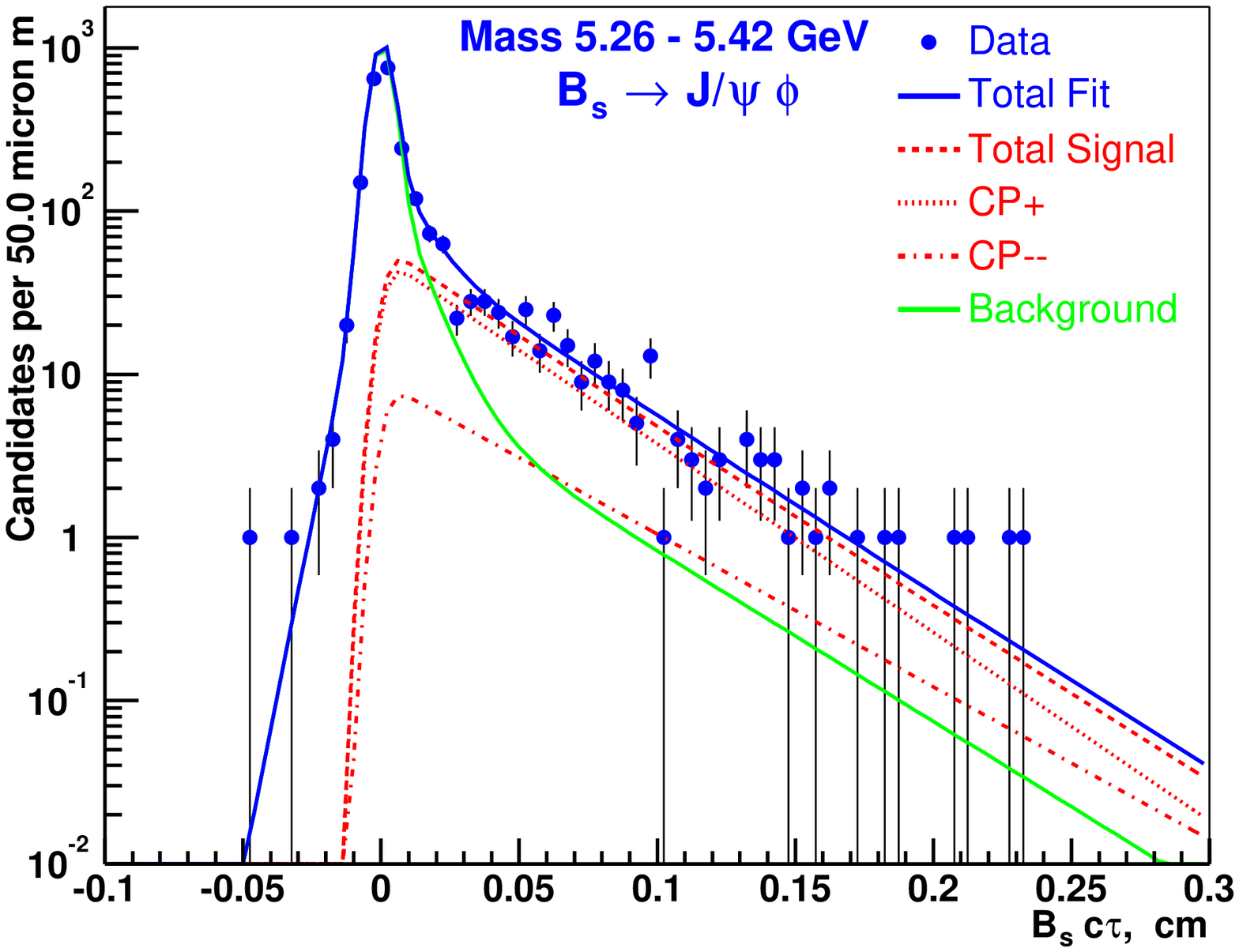 ,height=2in}
\end{tabular}
\caption{Mass (a) and lifetime (b) distributions for the D\O\ $\Delta \Gamma$ analysis.
 \label{fig:dgammamass}}
\end{figure}

D\O\ integrates over $\phi$ and $\psi$ using efficiencies derived from Monte Carlo.  
The "transversity" angle, $\theta$, which is defined in the rest frame of the 
J/$\psi$ as the angle of the $\mu^+$ with respect to the axis perpendicular to the decay plane 
of the $\phi \to $ K$^+$K$^-$, is used to extract the CP information.  
Integration of the full angular distribution over $\phi$ and $\psi$ assuming flat acceptance demonstrates 
the relation between the amplitudes and transversity:
\begin{displaymath}
\frac{d\Gamma}{d\cos\theta} \propto \frac{3}{8}(\left| A_0(t) \right|^2 + \left| A_\|(t) \right|^2)
(1+\cos^2\theta)+
\frac{3}{4} \left| A_{\bot}(t) \right|^2  \sin^2\theta \nonumber .
\end{displaymath}
D\O\ performs a 19 parameter fit, similar to the lifetime fits discussed above but including 
additional parameters for R$_\bot (=A_\bot(0))$ and $\Delta \Gamma/ \overline{\Gamma}$. Mass and lifetime distributions
are shown in Figure~\ref{fig:dgammamass}. The result~\cite{D04557} of this fit is:
\begin{eqnarray*}
\overline{\tau}(B_s) & = & 1.30^{+0.13}_{-0.14} \pm 0.08 \mbox{ps} , \\
\Delta \Gamma / \overline{\Gamma} & = & 0.21^{+0.27}_{-0.40} \pm 0.20 , \\
R_\bot & = & 0.17 \pm 0.10 \pm 0.02 .
\end{eqnarray*}

This result can be improved by including measurements of the average flavor-specific
$B_s$ lifetime in semileptonic decays. The flavor-specific width is the average of 
the CP even and odd values. The fit to a single lifetime in semileptonic analyses 
results in a measured lifetime related to the mean of the CP even and odd values of:
$ 1/\tau_{fs} = \Gamma_{fs} = \overline{\Gamma} - (\Delta \Gamma)^2 / 2 \overline{\Gamma} 
+ O((\Delta \Gamma)^3 / \overline{\Gamma}^2)$.
 A fit using $\overline{\tau}_{fs} = 1.43 \pm 0.05$ ps results 
in an improved value of $\Delta \Gamma/ \overline{\Gamma} = 0.23^{+0.16}_{-0.17}$.  

\begin{figure}
\begin{center}
\epsfig{figure=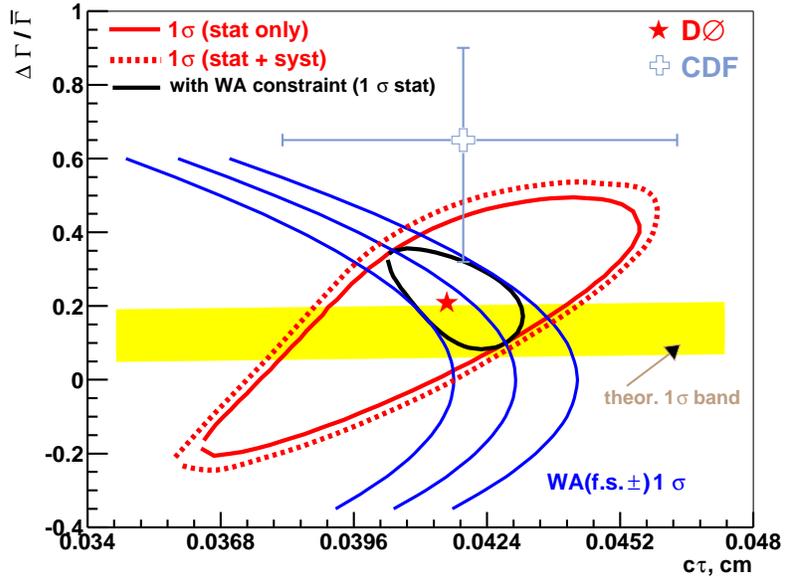 ,height=3in}
\end{center}
\caption{One $\sigma$ contours for $\Delta \Gamma$ plotted against fitted lifetime.  
Shown are the CDF (cross) and D\O\ (star) results, the theoretical band, and the 
world average lifetime constraint.\label{fig:bsdgamma}}
\end{figure}

This result 
can also be used to provide information on the CP violating phase, $\phi_s$, using the relation 
$\Delta \Gamma / \Gamma_{measured} = \Delta \Gamma / \Gamma _{CP conserving} \cos ^2 ( \phi_s )$
and assuming the predicted value of $\Delta \Gamma$/$\Gamma_{CP conserving}=0.12 \pm 0.05$.
The fitted value is $\left|\cos\phi_s \right| = 1.46^{+0.73}_{-0.69}$. A summary of the results is shown 
in Figure~\ref{fig:bsdgamma}.
 
\section{ Prospects}
Both D\O\ and CDF have upgrades planned that will allow them 
to effectively trigger on and reconstruct events at the highest 
planned Tevatron luminosities.  Tracking triggers for both experiments, 
which are compromised by occupancy at high luminosity, will increase in 
granularity.  
A new inner layer silicon detector for D\O\, similar to CDF's layer 00,
located 1.6 cm from the beam, will be installed 
in the Fall of 2005. D\O\ has also proposed a DAQ rate upgrade, which will 
allow the experiment to increase the rate to tape from $50$ to $100$ Hz, with 
the bulk of this increase dedicated to B physics.  

The Tevatron experiments have collected and analysed approximately $10 \% $ of the data 
set expected in Run II.  With an expected order of magnitude more data, 
lifetimes should be measured with $1 \%$ precision.  $\Delta \Gamma$ should be measured to 
a few percent, and states such as $\phi \phi$ and $\phi K^*$ will be included in 
the data sample.  With these data, experiments will address the detailed lifetime predictions
of HQET, have sensitivity to the CP phase $\phi_s$, continue the search for rare decays, 
and extend $B_s$ mixing measurements to the full range predicted by the standard model.

\end{document}